# COMPLEX BEHAVIOUR IN A DISCRETE COUPLED LOGISTIC MODEL FOR THE SYMBIOTIC INTERACTION OF TWO SPECIES

**Ricardo López-Ruiz** [*]
**Danièle Fournier-Prunaret** [#]

[*] Department of Computer Science and BIFI,
Facultad de Ciencias-Edificio B,
Universidad de Zaragoza,
50009 - Zaragoza (Spain).

[#] Institut National des Sciences Appliquées,
Systèmes Dynamiques (SYD), L.E.S.I.A.,
Avenue de Rangueil, 31077 Toulouse Cedex (France).

## Abstract

A symmetrical cubic discrete coupled logistic equation is proposed to model the symbiotic interaction of two isolated species. The coupling depends on the population size of both species and on a positive constant $\lambda$, named the mutual benefit. Different dynamical regimes are obtained when the mutual benefit is modified. For small $\lambda$, the species become extinct. For increasing $\lambda$, the system stabilizes in a synchronized state or oscillates in a 2 periodic orbit. For the greatest permitted values of $\lambda$, the dynamics evolves into a quasiperiodic, into a chaotic scenario or into extinction. The basins for these regimes are visualized as coloured figures on the plane. These patterns suffer different change as consequence of basins' bifurcations. The use of the critical curves let us to determine the influence of the zones with different number of first rank preimages in those bifurcation mechanisms.

***Keywords*** : *symbiotic species; population dynamics; coupled logistic maps; synchronization; complex patterns; invariant sets; critical curves; basins.*

# 1. DYNAMICS OF ISOLATED SPECIES: THE LOGISTIC MODEL

Imagine an island without any contact with the exterior. Living species there have no possibility to migrate looking for a new land with affordable resources. Thus, for instance, if the island has initially as inhabitants a couple of rabbits, they will reproduce exponentially. This expansion regime of the rabbits will finish for colonizing the whole island in a few generations. Hence, the island will become overpopulated. A new dynamical regime will be present now with a natural population control mechanism due to the overcrowding.

If $x_n$ represents the population after $n$ generations, let us suppose this variable bounded in the range $0 < x_n < 1$. The *activation or expanding phase* is controlled by the term $\mu x_n$ proportional to the current population $x_n$ and to the constant *growth rate* $\mu$. Resource limitations bring the system to an *inhibition or contracting phase* directly related with overpopulation. The term $(1 - x_n)$ can denote how far the system is of overcrowding. Therefore, if we take the product of both terms as the most simple approach to the population dynamics, the model

$$x_{n+1} = \mu x_n (1 - x_n)$$

gives account of its evolution. This is the so-called *logistic map*, where $0 < \mu < 4$ in order to assure $0 < x_n < 1$. The continuous version of this model was originally introduced by Verhulst [Verhulst, 1845] in the nineteenth century and it has been a subject of study in the last century as a tool to be applied to the most diverse phenomenology [May, 1976] or as an object interesting to analyze by itself from a mathematical point of view [Collet & Eckmann, 1980 ; Mira, 1987].

The dynamical behavior of the logistic equation when the growth rate is modified is as follows:

(i) $0 < \mu < 1$: The growth rate is not big enough to stabilize the population. It will drop and the specie will become extinct.
(ii) $1 < \mu < 3$: A drastic change is obtained when $\mu$ is greater than 1. The non vanishing equilibrium between the two competing forces, reproduction on one hand and resource limitation on the other, is now possible. The population reaches, independently of its initial conditions, a fixed value that is maintained in time.
(iii) $3 < \mu < 3.57$: A cascade of sudden changes provokes that the population oscillates in cycles of period $2^n$, where $n$ increases from 1, when $\mu$ is close to 3, to infinity when $\mu$ is approaching the critical value 3.57. This is named the period-doubling cascade.
(iv) $3.57 < \mu < 3.82$: When the parameter moves, the system alternates between periodical behaviours with high periods on parameter interval windows and *chaotic regimes* for parameter values not located in intervals. The population can be not predictable although the system is deterministic. The chaotic regimes are observed for a given value of μ on sub-intervals of [0,1].
(v) $3.82 < \mu < 3.85$: The orbit of period 3 appears for $\mu = 3.82$ after a regime where unpredictable bursts, named *intermittences*, have become rarer until their disappearance in the three-periodic time signal. The existence of the period 3 orbit means, such as the Sarkovskii theorem tells us, that all periods are possible for the population dynamics, although, in this case, they are not observable due to their instability. What it is observed in this range is the period-doubling cascade $3 * 2^n$.

- (vi) $3.85 < \mu < 4$: Chaotic behaviour with periodic windows is observed in this interval.
- (vii) $\mu = 4$: The chaotic regime is obtained on the whole interval $[0,1]$. This specific regime produces dynamics, which looks like random. The dynamics has lost near-all its determinism and the population evolves as a random number generator.

Therefore, there are essentially three remarkable dynamical behaviours in this system: the period doubling route to chaos around the value $\mu \approx 3.57$ [Feigenbaum, 1978], the time signal complexification by intermittence in the neighbourhood of $\mu \approx 3.82$ [Pomeau & Manneville, 1980] and the random-like dynamics for $\mu = 4$.

## 2. DYNAMICS OF TWO ISOLATED SYMBIOTIC SPECIES: A COUPLED LOGISTIC MODEL

Let us suppose now, under a similar scheme of expansion/contraction, that two symbiotic species $(x_n, y_n)$ are living on the island. Each one of them evolves following a logistic-type dynamics,

$$x_{n+1} = \mu(y_n) x_n (1 - x_n),$$
$$y_{n+1} = \mu(x_n) y_n (1 - y_n).$$

In this model, the symbiotic interaction between both species provokes that the growth rate $\mu(z)$ is varying with time. In fact, it depends of the population size of the *others* and of a positive constant $\lambda$ that we call the *mutual benefit*. As it is seen in the equations, we are thinking in a symmetrical interaction. Concretely, the particular dynamics of each one of the species is a logistic map whose parameter $\mu_n$ is not fixed, $x_{n+1} = \mu_n x_n (1 - x_n)$, but itself is forced to remain in the interval $[1,4]$. The existence of a nontrivial fixed point at each step $n$ ensures the nontrivial evolution of the system [López-Ruiz, 1991]. The simplest election for this growth rate is a linear function expanding all the interval $[1,4]$:

$$\mu(z) = \lambda(3z + 1),$$

with the mutual benefit $\lambda$ being a positive constant that the study has discovered to have sense in the range $0 < \lambda < 1.084$. Then, the model obtained to mimic the dynamics of two isolated symbiotic species takes the form:

$$x_{n+1} = \lambda(3y_n + 1) x_n (1 - x_n), \qquad (1)$$
$$y_{n+1} = \lambda(3x_n + 1) y_n (1 - y_n).$$

This application can be represented by $T_\lambda : [0,1] \times [0,1] \to \Re^2$, $T_\lambda(x_n, y_n) = (x_{n+1}, y_{n+1})$, where $\lambda$ is a real and adjustable parameter. In the following we shall write $T$ instead of $T_\lambda$ as the dependence on the parameter $\lambda$ is understood. Let us observe that when $y_n = 0$ or $x_n = 0$, the logistic dynamics for one isolated specie is recovered. In this case the parameter $\lambda$ takes the role of the parameter $\mu$.

At this point we must comment that the different choices of $\mu_n$ give a wide variety of dynamical behaviours. For instance, the application of this idea produces the on-off intermittence phenomenon when $\mu_n$ is chosen random [Platt *et al.*, 1993] or the adaptation to the edge of chaos when $\mu_n$ is a constant with a small time perturbation [Melby *et al.*, 2000]. Other systems built under this mechanism are models (a), (b) and (c) presented in [López-Ruiz, 1991 ; López-Ruiz & Pérez-Garcia, 1991]. Equations (1) correspond to model (a) of that work. Model (b) has been studied in detail in [López-Ruiz & Fournier-Prunaret , 2003], and a similar investigation of model (c) is presented in [Fournier-Prunaret & López-Ruiz, 2003].

In Sections 3 and 4, we study more accurately the model (1) from a dynamical point of view. In order to summarize, we explain first the dynamical behaviour of the coupled logistic system (1). When $\lambda$ is modified, it is as follows:

(i) $0 < \lambda < 0.75$: The mutual benefit is not big enough to allow a stable coexistence of both species and they will disappear.
(ii) $0.75 < \lambda < 0.866$: A sudden change is obtained when $\lambda$ is greater than 0.75. Both populations are synchronized to a stable non-vanishing fixed quantity when the initial populations overcome certain critical values. If the initial species are under these limits both will become extinct
(iii) $0.866 < \lambda < 0.957$: The system is now bi-stable. Each one of the species oscillates out-of-phase between the same two fixed values. This is a lag-synchronized state or in other words a stable 2-period orbit. There is still in this range the possibility of extinction when the initial populations are very small or close to the overcrowding.
(iv) $0.95 < \lambda < 1.03$: The system is not anymore on a periodic orbit. It acquires a new frequency and the dynamics is now quasiperiodic. Both populations oscillate among slightly infinitely many different states. Synchronization is lost. There are in this regime periodic windows where the system becomes newly lag-synchronized. Also, for initial populations nearly zero and for $\lambda < 1$, the species can not survive.
(v) $1.03 < \lambda < 1.0843$: The system is now in a chaotic regime. It is characterized by a noisy-like small oscillation around a synchronized state with non-periodic unpredictable bursts. Periodic oscillations can be also obtained for some particular values of the mutual benefit. Some other initial conditions are not meaningful or interpretable in this scheme because the system is going outward the square $[0,1] \times [0,1]$ and evolves toward infinity. The system 'crashes'. This sudden 'damage' is interpreted as some kind of catastrophe provoking the extinction of species.

Let us remark that although the equations are formed by logistic-type components, the logistic effects have been lost and a completely new scenario emerges when they are coupled. In this case, the symbiotic interaction makes the species to reach different stable states. Depending on the mutual benefit, the system can reach the extinction, a fixed synchronized state, a bi-stable lag-synchronized configuration, an oscillating dynamics among infinitely many possible states or a chaotic regime. We must highlight in this model the phenomenon of synchronization in the periodic regime [Boccaletti *et al.*, 2002], the transition to chaos by the Ruelle-Takens route [Eckmann, 1981] and the bursting events around a noisy-like synchronized state in the chaotic regime [Heagy *et al.*, 1995]. All these behaviours are caused by the symbiotic coupling of the species and are not predictable from the properties of the individual logistic evolution of each one of them. Moreover, this interaction implies a mutual profit for both species. In fact, when $\mu < 1$ one of the isolated species is extinct, but it can survive for $\lambda < 1$ if a small quantity of the other species is aggregated to the island. Hence, the symbiosis appears to be well held in this cubic model.

## 3. STABLE ATTRACTORS: SYMMETRY AND BIFURCATIONS

For the sake of clarity, firstly, we summarise the dynamical behaviour of model (1) when the mutual benefit $\lambda$ is inside the interval $0 < \lambda < 1.0843$. The different parameter regions where the mapping $T$ has stable attractors are given in the next table.

| INTERVAL | NUMBER OF ATTRACTORS | ATTRACTORS |
|---|---|---|
| $0 < \lambda < 0.75$ | 1 | $p_0$ |
| $0.75 < \lambda < 0.866$ | 2 | $p_0$, $p_4$ |
| $0.866 < \lambda < 0.957$ | 2 | $p_0$, $p_{5,6}$ |
| $0.957 < \lambda < 1$ | 2 | $p_0$, pair of invariant closed curves |
| $1 < \lambda < 1.03$ | 1 | pair of invariant closed curves |
| $1.03 < \lambda < 1.032$ | 1 | pair of weakly chaotic rings |
| $1.032 < \lambda < 1.0843$ | 1 | symmetric chaotic attractor (or frequency lockings) |

The meaning of all these attractors is explained in the next sections.

### 3.1 *Symmetry*

This model has reflection symmetry $P$ through the diagonal $\Delta = \{(x,x), x \in R\}$. If $P(x,y) = (y,x)$ then $T$ commutes with $P$:

$$T[P(x,y)] = P[T(x,y)].$$

Note that the diagonal is $T$-invariant, $T(\Delta) = \Delta$. In general, if $\Omega$ is an invariant set of $T$, $T(\Omega) = \Omega$, so is $P(\Omega)$ due to the commutation property: $T[P(\Omega)] = P[T(\Omega)] = P(\Omega)$. It means that if $\{p_i, i \in N\}$ is an orbit of $T$, so is $\{P(p_i), i \in N\}$. In fact, if some bifurcation happens in the half plane below the diagonal so it is in the above half plane, and vice versa. The dynamical properties of the two halves of phase space separated by the diagonal are interconnected by the symmetry.

Also if the set $\Gamma$ verifies $P(\Gamma) = \Gamma$, so is $T(\Gamma)$. Then the $T$-iteration of a reflection symmetrical set continues to keep the reflection symmetry through the diagonal. It is worth to note that the square $[0,1] \times [0,1]$ is invariant for µ<1, but not anymore for µ>1.

### 3.2 *Fixed Points, 2-Cycles and Closed Invariant Curves*

We focus our attention on bifurcations playing an important role in the dynamics, those happening in the interval $0 < \lambda < 1.0843$. In this range, there exist stable attractors for each value of $\lambda$ and it will have sense to study their basins of attraction, that is, the initial populations leading to the each one of the existing final asymptotic configurations.

The restriction of $T$ to the diagonal is a one-dimensional cubic map, which is given by the equation $x_{n+1} = \lambda(3x_n + 1)x_n(1 - x_n)$. The restriction of the map $T$ to the axes reduces to the logistic map $x_{n+1} = f(x_n)$ with $f(x) = \lambda x(1 - x)$. Thus the solutions of $x_{n+1} = x_n$ are the fixed points $p_0, p_3, p_4$ on the diagonal and $p_1, p_2$ on the axes:

$$p_0 = (0,0),$$

$$p_1 = \left(\frac{\lambda - 1}{\lambda}, 0\right), p_2 = \left(0, \frac{\lambda - 1}{\lambda}\right),$$

$$p_3 = \frac{1}{3}\left\{1 - \left(4 - \frac{3}{\lambda}\right)^{\frac{1}{2}}, 1 - \left(4 - \frac{3}{\lambda}\right)^{\frac{1}{2}}\right\},$$

$$p_4 = \frac{1}{3}\left\{1 + \left(4 - \frac{3}{\lambda}\right)^{\frac{1}{2}}, 1 + \left(4 - \frac{3}{\lambda}\right)^{\frac{1}{2}}\right\}.$$

For $0 < \lambda < 1$, $p_0$ is an attractive node. For all the rest of parameter values, $p_0$ is a repelling node. The points $(p_1, p_2)$ exist for every parameter value and they are unstable for every value of $\lambda$.
For $0 < \lambda < 0.75$, $p_{3,4}$ are not possible solutions. When $\lambda = 0.75$ a saddle-node bifurcation on the diagonal generates $p_{3,4}$. For $0.75 < \lambda < 0.866$, $p_3$ is a saddle point and $p_4$ is an attractive node. In this parameter interval, the whole diagonal segment between $p_3$ and $p_4$ is locus of points belonging to heteroclinic trajectories connecting the two fixed points.
The point $p_4$ suffers a flip bifurcation when $\lambda = \sqrt{3}/2 \cong 0.866$. It generates a stable period-two orbit $p_{5,6}$ outside the diagonal. These points are obtained by solving the quadratic equation $\lambda(4\lambda + 3)x^2 - 4\lambda(\lambda + 1)x + 1 + \lambda = 0$. The solutions are:

$$p_5 = \left(\frac{2\lambda(\lambda + 1) + \sqrt{\lambda(\lambda + 1)(4\lambda^2 - 3)}}{\lambda(4\lambda + 3)}, \frac{2\lambda(\lambda + 1) - \sqrt{\lambda(\lambda + 1)(4\lambda^2 - 3)}}{\lambda(4\lambda + 3)}\right),$$

$$p_6 = \left(\frac{2\lambda(\lambda + 1) - \sqrt{\lambda(\lambda + 1)(4\lambda^2 - 3)}}{\lambda(4\lambda + 3)}, \frac{2\lambda(\lambda + 1) + \sqrt{\lambda(\lambda + 1)(4\lambda^2 - 3)}}{\lambda(4\lambda + 3)}\right).$$

For $\lambda = 0.957$, these period-2 symmetric points lose stability via a Neimark-Hopf bifurcation. The set of points $p_{5,6}$ gives rise to a period 2 set of two stable closed invariant curves. The size of these symmetric invariant curves grow when $\lambda$ increases into the interval $0.957 < \lambda < 1$, and, for some values of $\lambda$, frequency locking windows are obtained.

The period two cycles on the axes appear by a period doubling bifurcation, and are found by solving the cubic equation: $\lambda^3 x^3 - 2\lambda^3 x^2 + (\lambda^3 + \lambda^2)x + 1 - \lambda^2 = 0$. They have existence for $\lambda > 3$. The solutions are:

$$p_7 = \left(\frac{(\lambda+1)-\sqrt{(\lambda+1)(\lambda-3)}}{2\lambda}, 0\right) \leftrightarrow p_8 = \left(\frac{(\lambda+1)+\sqrt{(\lambda+1)(\lambda-3)}}{2\lambda}, 0\right),$$

$$p_9 = \left(0, \frac{(\lambda+1)-\sqrt{(\lambda+1)(\lambda-3)}}{2\lambda}\right) \leftrightarrow p_{10} = \left(0, \frac{(\lambda+1)+\sqrt{(\lambda+1)(\lambda-3)}}{2\lambda}\right).$$

Observe that the restriction of the map $T$ to the axes is the logistic map, so that its dynamics gives rise to the well known cyclic logistic behaviour on the axes, as explained in section 1.

### *3.3 Transition to Chaos*

The two closed invariant curves approach the stable invariant set of the hyperbolic point $p_4$ on the diagonal for $\lambda$ slightly larger than 1 (Fig. 1a). At first sight, for $\lambda \approx 1.029$, the system still seems quasi-periodic but a finer analysis reveals the fingerprints of chaotic behaviour. Effectively, a folding process takes place in the two invariant sets, cf. [Frouzakis *et al.*, 1997], which gives rise to the phenomenon of weakly chaotic rings when the invariant set intersects itself (Fig. 1b) (cf. [Mira *& al.*, 1996] p529). For $\lambda \approx 1.032$, the tangential contact of the two symmetric invariant sets with the stable set of the saddle $p_4$ on the diagonal leads to the disappearance of those two weakly chaotic rings. Just after the contact, infinitely many repulsive cycles appear due to the creation of homoclinic points and a single and symmetric chaotic attractor appears (Fig. 2a). For $1.032 < \lambda < 1.0843$, this chaotic invariant set folds strongly around $p_4$, and the dynamics becomes very complex (Fig. 2b). When the limit value $\lambda = 1.084322$ is reached, the chaotic area becomes tangent to its basin boundary, the mapping iterates can escape to infinite and the attractor disappears by a contact bifurcation ([Mira *& al.*, 1987], chap. 5). The time behaviour of the system can be seen in Fig. 3(a-c).

### 4. BASIN FRACTALIZATION

Let us now see how the different initial populations evolve toward its asymptotic stable state. This is exactly the problem of considering the *basins* of the different attractors of model (1). For the sake of coherence, we consider the square $[0,1] \times [0,1]$ as the source of initial conditions having sense in our biological model, i.e., in the map $T$. Let us say at this point that basins constitute an interesting object of study by themselves. If a colour is given to the basin of each attractor, we obtain a coloured figure, which is a phase-plane visual representation of the asymptotic behaviour of the points of interest. The strong dependence on the parameters of this coloured figure generates a rich variety of complex patterns on the plane and gives rise to different types of basin fractalization. See, for instance, the work done by Gardini *et al.* [1994] and also by López-Ruiz & Fournier-Prunaret [2003] in this direction. It is now our objective to analyze the parameter dependence of basin fractalization of model (1) by using the technique of critical curves.

### *4.1 Definitions and General Properties of Basins and Critical Curves*

The set $D$ of initial conditions that converge towards an attractor at finite distance when the number of iterations of $T$ tends toward infinity is the basin of the attracting set at finite distance. When only one attractor exists at finite distance, $D$ is the basin of this attractor. When several attractors at finite distance exist, $D$ is the union of the basins of each attractor. The set $D$ is invariant under backward iteration $T^{-1}$ but not necessarily invariant by $T$: $T^{-1}(D) = D$ and $T(D) \subseteq D$. A basin may be connected or non-connected. A connected basin may be simply connected or multiply connected, which means connected with holes. A non-connected basin consists of a finite or infinite number of connected components, which may be simply or multiply connected. The closure of $D$ includes also the points of the boundary $\partial D$, whose sequences of images are also bounded and lay on the boundary itself. If we consider the points at infinite distance as an attractor, its basin $D_\infty$ is the complement of the closure of $D$. When $D$ is multiply connected, $D_\infty$ is non-connected, the holes (called lakes) of $D$ being the non-connected parts (islands) of $D_\infty$. Inversely, non-connected parts (islands) of $D$ are holes of $D_\infty$ [Mira *et al.*, 1996].

In Sec. 3, we explained that the map (1) may possess one or two attractors at a finite distance. The points at infinity constitute the third attractor of $T$. Thus, if a different colour for each different basin is given we obtain a coloured pattern in the square $[0,1] \times [0,1]$ with a maximum of two colours. In the present case, the phenomena of finite basins disappearance have their origin in the competition between the attractor at infinity (whose basin is $D_\infty$) and the attractors at finite distance (whose basin is $D$). When a bifurcation of $D$ takes place, some important changes appear in the coloured figure, and, although the dynamical causes cannot be clear, the coloured pattern becomes an important visual tool to analyze the behaviour of basins.

Critical curves are an important tool used to study basin bifurcations. They were introduced by Mira in 1964 (see [Mira *et al.*, 1996] for further details). The map *T* is said to be noninvertible if there exist points in state space that do not have a unique rank-one preimage under the map. Thus the state space is divided into regions, named $Z_i$, in which points have *i* rank-one preimages under *T*. These regions are separated by the so called critical curves $LC$, which are the images of the curves $LC_{-1}$. If the map *T* is continuous and differentiable, $LC_{-1}$ is the locus of points where the determinant of the Jacobean matrix of *T* vanishes. When initial conditions are chosen to both sides of $LC$, the rank-one preimages appear or disappear in pairs. (See also the glossary for different technical terms used along this work).

### *4.2 Critical Curves and $Z_i$ – Regions of T*

In our case, the map $T$ defined in (1) is noninvertible. It has a non-unique inverse. As we know, $LC = T(LC_{-1})$. $LC_{-1}$ is the curve verifying $|DT(x,y)| = 0$, where $DT(x,y)$ is the Jacobean matrix of $T$. It is formed by the points $(x,y)$ that satisfy the equation:

$$27x^2 y^2 + 3x^2 y + 3xy^2 - 6x^2 - 6y^2 - 8xy + x + y + 1 = 0. \qquad (2)$$

Hence, $LC_{-1}$ is independent of $\lambda$ parameter and is quadratic in $x$ and $y$. It can be seen that $LC_{-1}$ is a curve of four branches, with two horizontal and two vertical asymptotes. The branches $LC_{-1}^{(1)}$ and $LC_{-1}^{(2)}$ have as horizontal asymptote the line $y = 0.419$ and the vertical asymptote in $x = 0.419$. The other two branches, $LC_{-1}^{(3)}$ and $LC_{-1}^{(4)}$, have the horizontal asymptote in $y = -0.530$ and the vertical one is the line $x = -0.530$. The values $0.419$ and $-0.530$ are the roots of the polynomial factor, $27x^2 + 3x - 6$, that multiplies the term $y^2$ in Eq. (2). It follows that the critical curve of rank-1, $LC^{(i)} = T(LC_{-1}^{(i)})$, $i = 1,2,3,4$, consists of four branches. The shape of $LC$ and $LC_{-1}$ is shown in Figs. 4(a-b). $LC$ depends on $\lambda$ and separates the plane into three regions that are locus of points having 1, 3 or 5 distinct preimages of rank-1. They are named by $Z_i$, $i = 1,3,5$, respectively (Figure 4b). Observe that the set of points with three preimages of rank-1, $Z_3$, is not connected and is formed by five disconnected zones in the plane. Let us remark that $LC_{-1}$ has the reflection symmetry through the diagonal: $P(LC_{-1}) = LC_{-1}$. Then every critical curve of rank-(k+1), $LC_k = T^{k+1}(LC_{-1})$, will conserve this symmetry: $P(LC_k) = LC_k$.

We see in Fig. 4(b) that the four branched $LC$-curve divides the diagonal $\Delta$ in five intervals. If we know the number $i$ of preimages of rank-1 of each segment on the diagonal, the number of preimages of rank-1 of each $Z_i$-zone of the plane is also determined. This calculation has been performed in [López-Ruiz & Fournier-Prunaret, 2003]. The number of rank-1 preimages of a point $(x', x')$ on the diagonal can be summarised in the following table:

| INTERVAL | $x' < x'_{2d}$ | $x'_{2d} < x' < x'_{2h}$ | $x'_{2h} < x' < x'_{1d}$ | $x'_{1d} < x' < x'_{1h}$ | $x' > x'_{1h}$ |
|---|---|---|---|---|---|
| NUMBER OF PREIMAGES | 3 | 5 | 3 | 1 | 3 |

The coordinates of the points marking the frontier between the different $Z_i$-zones on the diagonal are: $x'_{1d} \approx 0.65\lambda$ and $x'_{2d} \approx -0.1\lambda$, $x'_{1h} = 4\lambda$ and $x'_{2h} \approx 0.44\lambda$. For example, the origin $p_0$ is always in the $Z_5$-zone. It is located into the interval limited by $x'_{2d}$ and $x'_{2h}$. In fact, its preimages are $(1,1), (-1/3, -1/3)$ and $p_0$ itself on the diagonal, and $(1,0), (0,1)$ out of the diagonal. According to the nomenclature established in Mira *et al.* [1996], the map (1) is of type $Z_3 - Z_5 \succ Z_3 - Z_1 \prec Z_3$.

### 4.3 *Types of Basins in T*

Depending on $\lambda$, three different types of patterns are obtained in the square $[0,1] \times [0,1]$. We proceed to present them and to explain the role played by critical curves in the bifurcations giving rise to the third basin type.

### 4.3.1 *One-Colour Pattern: Extinction of Species,* $0 < \lambda < 0.75$

In this regime, any given initial population evolves toward the extinction. The mutual benefit is small to allow the surviving of the species. Then, all initial conditions tend to zero under iteration of $T$. A pattern of only one colour is obtained (Fig. 5a).

If we regard the behaviour of $T$ in the whole plane $R^2$, $D$ undergoes an interesting bifurcation consisting in the transition from a connected to a non-connected basin (Fig. 5b). It takes place when $\lambda$ increases from $\lambda \approx 0.39$ to $\lambda \approx 0.61$. When $D$ becomes non-connected, it is made up of the immediate basin $D_0$ containing the single attractor $(p_0)$ and infinitely many small regions without connection (islands). This disaggregation is the result of infinitely many contact bifurcations, which are explained in [López-Ruiz & Fournier-Prunaret, 2003]. Such phenomenon can be also found in some quadratic $Z_0 - Z_2$ maps [Mira & Rauzy, 1995].

### *4.3.2 Two-Colour Pattern: Extinction or non-trivial Evolution of Species,* $0.75 < \lambda < 1$.

A sudden change suffers the basin for $\lambda = 0.75$. A second attractor $p_4$ appears and a ball of initial conditions around this point is attracted by this synchronized state. The coexistence between both species can reach a non-null stable value in this regime when $0.75 < \lambda < 0.866$. All the rest of initial conditions on the square $[0,1] \times [0,1]$ continue to shrink to the origin, then to be extinct. The basin is a two-colour pattern (Fig. 6a). When $0.866 < \lambda < 0.957$, $p_4$ bifurcates to a 2-periodic orbit and the system becomes now a lag-synchronized oscillation. The colour corresponding to this last state has gained space on the zero state in the two-colour pattern (Fig. 6b). When $0.957 < \lambda < 1$, synchronization is finally lost and the system becomes quasiperiodic. Only the corners of the square $[0,1] \times [0,1]$ lead to extinction in the two-colour pattern (Fig. 6c).

If we regard the total basin in $R^2$, $D$ seems to be formed by the square $D_0 \equiv [0,1] \times [0,1]$, which contains the attracting set at finite distance, and four small like-triangled regions linked to the square by four narrow arms. These arms shrink when $\lambda$ approaches 1, and disappear for $\lambda = 1$ when the origin $p_0$ undergoes a transcritical bifurcation. The main part of $D$ is then a disconnected pattern of five components: the square $D_0$, a triangle-shaped component located in a $Z_3$ neighbourhood of the vertex point $(-1/3, -1/3)$ (preimage of rank-1 of the point $p_0$), and the three triangle-shaped regions that are preimages of rank-1 of the latter component (Fig. 6d).

### *4.3.3 Two-Colour Pattern: non-trivial Evolution or Catastrophe of Species,* $1 < \lambda < 1.0843$

The extinction by exhaustion of the resources is not possible now. But a new phenomenon takes place in this range of the parameter. Some initial conditions can give rise to an evolution that surpass the boundaries of the square $[0,1] \times [0,1]$ and tend to infinity. We interpret this behaviour as some kind of internal catastrophe (war, epidemics, etc.) leading also to extinction. Although we are aware of its disconcerting meaning, this would imply that an internal catastrophe can follow in this model as a consequence of the population start from some particular initial conditions. All the rest of initial conditions bring the system to a quasiperiodic state when $1 < \lambda < 1.03$ or a chaotic dynamical regime when $1.03 < \lambda < 1.0843$ (Fig. 7(a-f)). Therefore, a two-colour basin is also obtained in this range of $\lambda$ parameter.

In a more detailed form, the basin bifurcations happen as follows. Points $(1,0)$ and $(0,1)$ cross through $LC^{(2)}$ when $\lambda = 1$. When $\lambda > 1$, it makes appear two regions $S_1^1$ and $S_2^1$ inside $[0,1] \times [0,1]$, which are part of $D_\infty$ and are located in a $Z_3$ zone (Fig. 7a). The square is not anymore invariant by T. The rank-one preimages of $S_1^1$ and $S_2^1$, respectively $S_1^{-1}$ and $S_2^{-1}$ are two new semicircular regions and intersect $LC_{-1}^{(2)}$. They are located in the vicinity of points $(1,0.5)$ and $(0.5,1)$, preimages of $(1,0)$ and $(0,1)$ (Fig. 7a-b). When $\lambda$ increases, the two semicircular shaped zones of $D_\infty$, $S_1^{-1}$ and $S_2^{-1}$, located in the immediate basin $D_0 \equiv [0,1] \times [0,1]$ in the neighbourhood of points $(1,0.5)$ and $(0.5,1)$, grow in size.

For $\lambda \approx 1.0801$, the basin undergoes a contact bifurcation. $D_\infty$ crosses through $LC^{(2)}$ and two bays (headlands of $D_\infty$), $H_{01}$ and $H_{02}$, are created in a $Z_3$ area (Fig. 7c). Their rank-1 preimages, $H_{01}^{(1)}$ and $H_{02}^{(1)}$, are holes (lakes) intersecting $LC_{-1}^{(2)}$ into the middle $Z_3$-region. Rank-1 preimages of the latter holes generate four new lakes in $D_0$, $H_{0i}^{(21)}$ and $H_{0i}^{(22)}$, $i = 1,2$. Preimages with increasing rank give rise to an arborescent sequence of lakes. The accumulation points of this infinite sequence of holes are the two unstable foci $p_{5,6}$ and their rank-1 preimages inside the basin. When $\lambda \approx 1.0806$, $H_{01}^{(21)}$ and $H_{02}^{(21)}$ cross through $LC^{(2)}$ (Fig 7d). This new contact bifurcation is the germ of a new arborescent and spiralling sequence of lakes converging towards the same accumulation points. When $\lambda$ increases values, new holes intersect $LC^{(2)}$ and give rise to new holes crossing through $LC_{-1}^{(2)}$ and new sequences of lakes converging towards the unstable foci $p_{5,6}$ and their preimages. Due to the fact that the preimages have a finite number of accumulation points, the structure is not fractal. A similar phenomenology has been found and studied in $Z_0 - Z_2$ maps [Mira *et al.*, 1994]. When $\lambda$ increases ($\lambda \approx 1.0835$), the chaotic attractor, which is limited by arcs of $LC_n$ curves, is destroyed by a contact bifurcation with its basin boundary (Fig. 7e). A new dynamical state arises. The infinite number of unstable cycles and their rank-n images belonging to the existing chaotic area before the bifurcation define a strange repulsor which manifests itself by chaotic transients (Fig. 7f). For $\lambda \approx 1.085$ the basin pattern disappears definitively.

## 5. CONCLUSIONS

One-dimensional and two-dimensional mappings are simple models that have been extensively studied as models of population dynamics [May, 1976 ; Kendall & Fox, 1998], as ingredients of other more complex systems [Kaneko, 1983 , Kapral, 1985] or as independent objects of interest [Mira, 1987]. Specifically, different two-dimensional coupled logistic maps are found in the literature of several fields, such as physics, engineering, biology, ecology and economics ([Yuan *et al.,* 1983], [Hogg & Huberman, 1984], [Van Biskirk & Jeffries, 1985], [Schult *et al.,* 1987], [De Sousa Vieira *et al.,* 1991]).

The models scattered in the literature on the three main types of population interaction, namely predator-prey situation, competition or mutualism among species, are usually stated as quadratic equations [Murray, 2002, and references therein]. In this work, we have reinterpreted a cubic two-dimensional coupled logistic equation [López-Ruiz, 1991] as a discrete model to explain the evolution of two symbiotically interacting species. The symbiotic interaction between both species is population-size dependent and is controlled by a positive constant $\lambda$ that we call the *mutual benefit*. Depending on $\lambda$, the system can reach extinction due to the small mutual benefit or the lack of resources, it can stabilize in a synchronized state or oscillates in a 2 periodic orbit for intermediate $\lambda$ or it can evolve in a quasiperiodic or chaotic regime for the greatest $\lambda$. In this last scenario, there are also initial conditions leading the system to extinction. This kind of extinction has been interpreted as an internal catastrophe caused, for instance, by political decisions or by a deficient health prevision system and not by the resources to become exhausted. Another remarkable property of the model is the fact that when $\mu < 1$ one of the isolated species is extinct, but it can survive for $\lambda < 1$ when it interacts symbiotically with one of the other species. Then, symbiosis seems to be well held in this model.

Different complex colour patterns on the plane have been obtained when the mutual benefit is modified. If $0 < \lambda < 0.75$, all the dynamics is attracted by the origin and a one-colour pattern is found. When $0.75 < \lambda < 1$ the dynamics can settle down in two possible attractors and the basins are now characterized by two colours. Finally, if $1 < \lambda < 1.0843$, the two-colour basins are consequence of the two possible asymptotic states: a quasiperiodic or chaotic finite distance attractor and an additional one located at infinity.

Critical curves have been used in order to understand the basin bifurcations found in this system. Hence, common features with those present in the simplest and well-studied case of $Z_0 - Z_2$ maps are now more evident for this map of $Z_3 - Z_5 \succ Z_3 - Z_1 \prec Z_3$ type. A detailed study of the different fractalization mechanisms for the whole range of $\lambda$ parameter for a similar coupled logistic equation was performed in [López-Ruiz & Fournier-Prunaret, 2003]. Let us remark that the rich dynamics and the complex patterns produced on the plane in this model are controlled by a single parameter, in this case, the mutual benefit between the interacting species.

**Acknowledgements**

We would like to thank Prof. Mira and Dr. Taha for their helpful comments. R. L.-R. wishes also to thank the *Systèmes Dynamiques* Group at INSA (Toulouse) for his kind hospitality. This work was supported by the Spanish MCYT project HF2002-0076 and French research project EGIDE-PICASSO 05125VC.

--------------------------

# GLOSSARY

INVARIANT: A subset of the plane is invariant under the iteration of a map if this subset is mapped exactly onto itself.

ATTRACTING: An invariant subset of the plane is attracting if it has a neighbourhood every point of which tends asymptotically to that subset or arrives there in a finite number of iterations.

CHAOTIC AREA: An invariant subset that exhibits chaotic dynamics. A typical trajectory fills this area densely.

CHAOTIC ATTRACTOR: A chaotic area, which is attracting.

BASIN: The basin of attraction of an attracting set is the set of all points, which converge towards the attracting set.

IMMEDIATE BASIN: The largest connected part of a basin containing the attracting set.

ISLAND: Non-connected region of a basin, which does not contain the attracting set.

LAKE: Hole of a multiply connected basin. Such a hole can be an island of the basin of another attracting set.

HEADLAND: Connected component of a basin bounded by a segment of a critical curve and a segment of the immediate basin boundary of another attracting set, the preimages of which are islands.

BAY: Region bounded by a segment of a critical curve and a segment of the basin boundary, the successive images of which generate holes in this basin, which becomes multiply connected.

CONTACT BIFURCATION: Bifurcation involving the contact between the boundaries of different regions. For instance, the contact between the boundary of a chaotic attractor and the boundary of its basin of attraction or the contact between a basin boundary and a critical curve $LC$ are examples of this kind of bifurcation.

# Figure Captions

**Fig. 1: (a)** Attractive closed invariant curves for $\lambda = 1.031$, and **(b)** Enlargement of (a), where weakly chaotic rings limited by segments of critical curves $LC_n$ can be observed.

**Fig. 2: (a)** Symmetric chaotic attractor for $\lambda = 1.0831$. **(b)** Complex folding process around $p_4$ for same value of $\lambda$.

**Fig. 3:** Asymptotic temporal behaviour of the dynamics: **(a)** for $\lambda = 0.9$, **(b)** for $\lambda = 1$, **(c)** for $\lambda = 1.08$.

**Fig. 4: (a)** Critical curves $LC_{-1}^{(i)}$, $i = 1,2,3,4$. **(b)** Critical curves $LC^{(i)}$, $i = 1,2,3,4$, for $\lambda = 0.4$. Observe the different $Z_j$-zones, $j = 1,3,5$.

**Fig. 5: (a)** One-colour basin for $\lambda = 0.45$. The only attractor is the origin. **(b)** Fractal pattern of islands when the whole $R^2$ plane is considered as a source of initial conditions.

**Fig. 6: (a)** Basin for $\lambda = 0.8$. The two colours correspond to the basins of the two existing attractors: the synchronized state on the diagonal and the origin. **(b)** For $\lambda = 0.9$, the central coloured ball in the square is the basin of a 2-periodic orbit. **(c)** For $\lambda = 0.98$, the central coloured area is the basin of two attractive closed invariant curves. **(d)** Pattern of the basin in the whole plane. It is formed by the square $[1,0] \times [0,1]$, which contains the attractors, and four small like-triangle regions linked to the square by four narrow arms for $\lambda = 0.9$.

**Fig. 7: (a-b)** Basin for $\lambda = 1.03$. One colour corresponds to basin of the attractive invariant curves and the other one to basin of infinity $D_\infty$. **(c)** For $\lambda = 1.0803$ first rank holes $H_{01}^{(1)}$ and $H_{02}^{(1)}$ (and higher rank preimages holes) of the bays $H_{01}$ and $H_{02}$, respectively. **(d)** New arborescent sequence of holes created from the crossing of $H_{01}^{(21)}$ and $H_{02}^{(21)}$ with $LC^{(2)}$ for $\lambda = 1.082$, **(e)** Chaotic attractor and its basin for $\lambda = 1.083$, **(f)** Chaotic transient for $\lambda = 1.0838$.